\documentclass{emulateapj}

\usepackage{apjfonts}
\usepackage{amsmath}
\usepackage{amssymb}
\usepackage{natbib}
\usepackage{xspace}
\usepackage{graphicx}
\usepackage{rotate} 
\usepackage{subfigure}
\usepackage{float}

\newcommand{\Msun}{\ensuremath{M_\odot}\xspace}

\newcommand{\nh}{$N_{\text{H}}$\xspace}

\newcommand{\xte}{\textsl{RXTE}\xspace}

\newcommand{\suzaku}{\textsl{Suzaku}\xspace}

\newcommand{\xmm}{\textsl{XMM-Newton}\xspace}
\newcommand{\chandra}{\textsl{Chandra}\xspace}

\newcommand{\cena}{Cen~A\xspace}

\newcommand{\etal}{et al.~}

\newcommand{\ttt}{$\times 10^{22}$\,cm$^{-2}$\xspace}

\shorttitle{Cen~A}
\shortauthors{Rivers et al.}

\begin{document}

\title{An Occultation Event in Centaurus~A and the Clumpy Torus Model} 

\author{Elizabeth~Rivers, Alex~Markowitz, Richard~Rothschild}
\affiliation{University of California, San Diego, Center for
  Astrophysics and Space Sciences, 9500 Gilman Dr., La Jolla, CA
  92093-0424, USA} 
\email{erivers@ucsd.edu}


\begin{abstract}

We have analyzed 16 months of sustained monitoring observations of \cena from the \textsl{Rossi X-ray Timing Explorer} to search
for changes in the absorbing column in the line of sight to the central nucleus.  We present time-resolved spectroscopy which indicates 
that a discrete clump of material transited the line of sight to the central illuminating source over the course of $\sim\,$170 days between 
2010 August and 2011 February with a maximum increase in the column density of about  8.4 $\times 10^{22}$\,cm$^{-2}$.
This is the best quality data of such an event that has ever been analyzed with the shape of the ingress and egress clearly seen.  
Modeling the clump of material as roughly spherical with a linearly decreasing density profile and assuming a distance from the central nucleus 
commensurate with the dusty torus we found that the clump would have a diameter of 1.4--2.4 $\times 10^{15}$\,cm with a central 
number density of n$_{\rm H}$ = 1.8--3.0 $\times 10^7$\,cm$^{-3}$.  This is consistent with previous results for a similar (though possibly 
much longer) occultation event inferred in this source in 2003--2004 and supports models of the molecular torus as a clumpy medium.

\end{abstract}



\section{Introduction}

Centaurus A (\cena hereafter) is one of the closest and consequently one of the brightest active galactic nuclei (AGN) in our sky.  
High resolution data from infrared and X-ray observatories (such as \textsl{Spitzer} and \chandra) have revealed many details about
the structure of its prominent jets as well as the core of the nucleus.  \cena has proved to be an excellent laboratory for studying the properties 
of blazar jets, however it also displays certain X-ray properties characteristic of a Seyfert galaxy, likely due to its orientation.  For example, X-ray 
spectroscopy shows a prominent Fe line at 6.4 keV originating in cool material far from the central nucleus (Evans \etal 2004) and strong 
absorption in the line of sight.  
Broad optical emission lines have never been detected in this source, marking it as a Seyfert 2.  A search for hidden broad lines through 
polarized scattered light by Alexander \etal (1999) ruled out the existence of a hidden broad line region (BLR).

Between the launch of the \textsl{Rossi X-ray Timing Explorer} (\xte) in 1996 and 2009 February, \cena was observed 13 times, 
each an extended exposure of $\sim\,$10--100 ks.  Rivers \etal (2011) performed spectral analysis on these data in the 3--100 keV band 
to measure longterm average spectral properties of \cena, such as the photon index ($\Gamma$ = 1.83$\pm0.01$) and the equivalent H 
column density (\nh = 16.9$\pm0.3 \times 10^{22}$\,cm$^{-2}$), and also confirming the lack of a Compton reflection hump in this source.  
Rothschild \etal (2011) analyzed the individual \cena observations, examining both spectral and temporal characteristics of this source.  
From these analyses it was discovered that for three observations between 2003 March and 2004 February, the column density of cold 
material along the line of sight to the nucleus increased by $\sim\,$60\%, from 16 $\times 10^{22}$\,cm$^{-2}$ to 26 $\times 10^{22}$\,cm$^{-2}$.  
From this it was inferred that a clump of material may have passed through the line of sight at a distance commensurate with the molecular torus.
Such an event is consistent with clumpy torus models such as that developed by Nenkova \etal (2008a; 2008b).
However there were only three data points in this interval and therefore the physical parameters of the inferred clump 
such as size, number density, and shape, were loosely constrained at best.

Similar short-term increases and decreases in \nh have been seen previously in a number of AGN, notably NGC~3227 (Lamer \etal 2003), 
NGC~1365 (Risaliti \etal 2009), NGC~4051 (Guainazzi \etal 1998), and H0557--385 (Longinotti \etal 2009).  
Most of these did not have the advantage of continuous monitoring over timescales of days to months.  The only exception to this is
Lamer \etal (2003), who attempted to fit a density profile to explain the smooth increase and decrease in \nh for NGC~3227 
over about 300 days as seen with \xte monitoring.  
Risaliti \etal (2009) analyzed a 60 ks \xmm observation of NGC~1365, finding very rapid changes in \nh on the order of only a few hours.  
From the short duration of these events they inferred that the material must be quite close to the central nucleus. 
Further analysis performed by Maiolino \etal (2010) with a 300 ks \suzaku observation confirmed these rapid variations and 
attributed them to comet-shaped BLR clouds transiting the line of sight on timescales of $\sim$\,50--100 ks. 

A sustained monitoring campaign of \cena with \xte began on 2010 January 1, with $\sim\,$1 ks snapshots every 2 days.
One goal of this observing campaign was to better quantify variation in the column density in this source and to search for
additional evidence of transits by discrete clumps of material, with the ability to place better constraints on their physical
characteristics.  To that end we have analyzed the \xte observations of \cena from the monitoring campaign, 
beginning 2010 January 1 up through 2011 April 20.  In this paper we present the results of this analysis 
with data reduction and analysis methods in presented in Section 2 and a discussion of our results in Section 3.

\section{Data Reduction and Analysis}

We analyzed 228 $\sim\,$1 ks snapshots of \cena with \xte's Proportional Counter Array (PCA; Jahoda \etal 2006).
For all PCA data extraction and analysis we used HEASOFT version 6.7 software.
Using standard extraction procedures (e.g. Rothschild \etal 2011) we extracted PCA STANDARD-2 data from PCU 2 only, 
using events from the top Xe layer only in order to maximize signal-to-noise.  
We applied standard screening criteria: $\geq$ 10$\degr$ from Earth's limb, $\geq$ 20 min since passage through
the peak of the South Atlantic Anomaly, satellite pointing $\leq$ 0.$\degr$01, and ELECTRON2 (particle flux) $\leq$ 0.1.
The background was estimated from the L7-240 background models for low source count rates 
($\lesssim$ 40 counts s$^{-1}$ PCU$^{-1}$) and from the Sky-VLE background model for higher count rates.

\begin{figure}
  \plotone{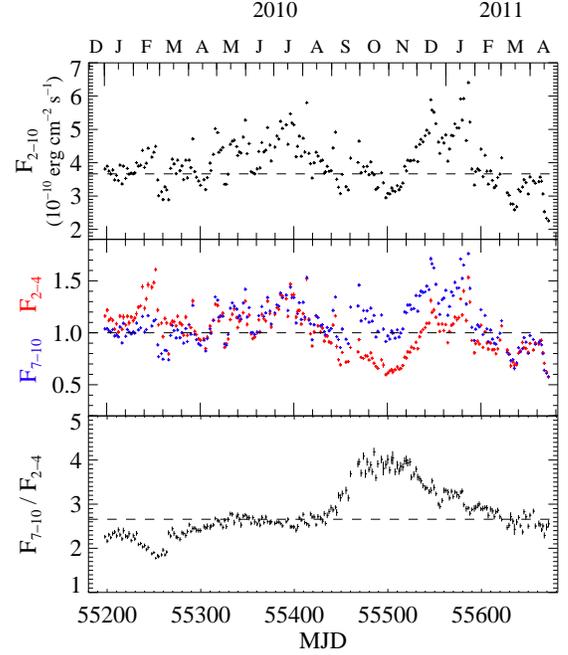}
  \caption{Light curves for the monitoring campaign from 2010 January 1 up through 2011 April 20.
  The top panel shows the 2--10 keV flux light curve.  The middle panel shows 2--4 and 7--10 keV fluxes normalized to their respective means.
  The bottom panel shows the F$_{7-10}$/F$_{2-4}$ ratio.}
  \label{figlc}
\end{figure}

The 2--10 keV flux light curve is shown in Figure \ref{figlc} along with fluxes in the 2--4 and 7--10 keV ranges.  
The ratio F$_{7-10}$/F$_{2-4}$ can indicate variations in \nh for Compton-thin sources which are independent of variations in the shape of
the power-law continuum.  There was a clear temporary increase 
in this ratio between MJD 55440 and 55600 that can be inferred to be a temporary increase in \nh caused by a cloud or clump
of material passing through the line of sight to the nucleus.  To confirm this we performed time-resolved spectral fitting.

All spectral fitting for this analysis was done using \textsc{xspec} version 12.5.1k with cross-sections from Verner et al.\ (1996) 
and solar abundances from Wilms et al.\ (2000).  Fitting clump geometries to the time resolved \nh values was performed in IDL version 6.3.
Uncertainties on spectral fit parameters were calculated using a point-to-point variance method (Vaghan \& Edelson 2001; Markowitz \etal 2003) 
at the 1$\sigma$ level for all parameters.  This method was appropriate for our analysis since the normal method tends to overestimate errors 
due to the background modeling for short observations such as ours.  

\begin{figure}
  \plotone{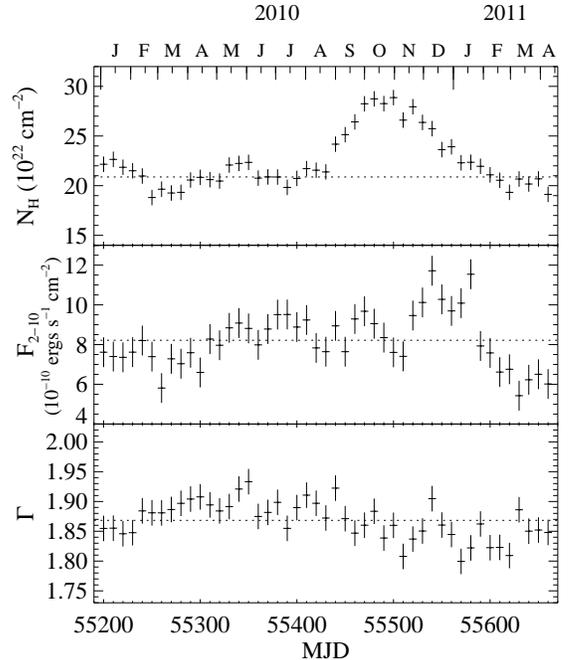}
  \caption{Time resolved spectral fitting parameters from 2010 January through 2011 April.  The top panel shows the column density along with
  a dotted line indicating the average value for the first 200 days of 20.8 $\times 10^{22}$\,cm$^{-2}$.  The middle panel shows the unabsorbed 
  2--10 keV flux, and the bottom panel the photon index, both with dotted lines indicating the average values for each.  
  Note that the photon index is consistent with maintaining a constant value of 1.87 $\pm 0.03$ throughout the monitoring. ~~~~~~~~~~~~~~~~~~~~~~~~~~~~~~~~~~~~~~~~~~~~~~~~~~~~~~~~~~~~~~~~~~~~~~~~~~~~~~~~~~~~}
  \label{fignh}
\end{figure}

\begin{figure}
  \plotone{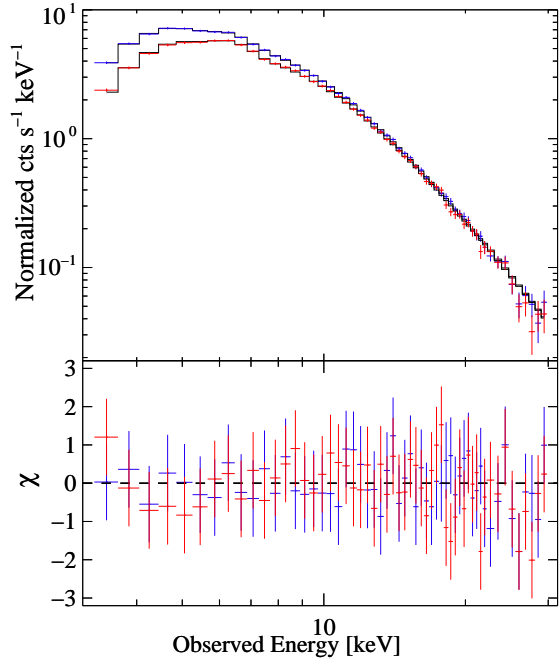}
  \caption{Two sample spectra with data and best-fit residuals.  Blue corresponds to an \nh of 20.7 $\times 10^{22}$\,cm$^{-2}$ 
  (MJD 55355--55365).  Red corresponds to an \nh of  28.9 $\times 10^{22}$\,cm$^{-2}$ (MJD 55495--55505).}
  \label{figspec}
\end{figure}

We analyzed PCA spectra in 10 day intervals with total exposures of 3--6 ks.  We included data from 3--30 keV in all time bins.
Our base model consisted of a power law with a fixed Galactic absorption column of 8.09 $\times 10^{20}$\,cm$^{-2}$ (Kalberla \etal 2005), 
an additional cold absorber with a free column density, and an Fe line modeled with a Gaussian.
Results from this fitting are shown in Figure \ref{fignh} and example spectra with models and best-fit residuals are shown in Figure
\ref{figspec}.  Reduced $\chi^2$ values were close to 1 in all cases.


It is clear that \nh increased significantly for $\sim\,$6 months between 2010 August and 
2011 February, rising from $\sim\,$20 $\times 10^{22}$\,cm$^{-2}$ to a maximum of 27 $\times 10^{22}$\,cm$^{-2}$.
The average column density in the 7 months preceding this event was 20.9 $\times 10^{22}$\,cm$^{-2}$.
The photon index values were consistent with a constant $\Gamma\,=\,1.87\pm 0.03$ through this 16 month period.  
Fe line parameters (which are not shown here) were poorly constrained but showed no evidence for strong variability.
We tried applying the \textsc{cabs} model to test whether the slight fluctuations in the
unabsorbed 2--10 keV flux were due to extra scattering by the increased amount of material in the line of sight.
This model is more commonly used for Compton-thick sources to model the attenuation of the power law by scattering,
however since the changes in \nh were less than $10^{23}$\,cm$^{-2}$, including this model did not affect the relative 
magnitude of the fluctuations and we did not use it in our final analysis.

\begin{figure}
  \plotone{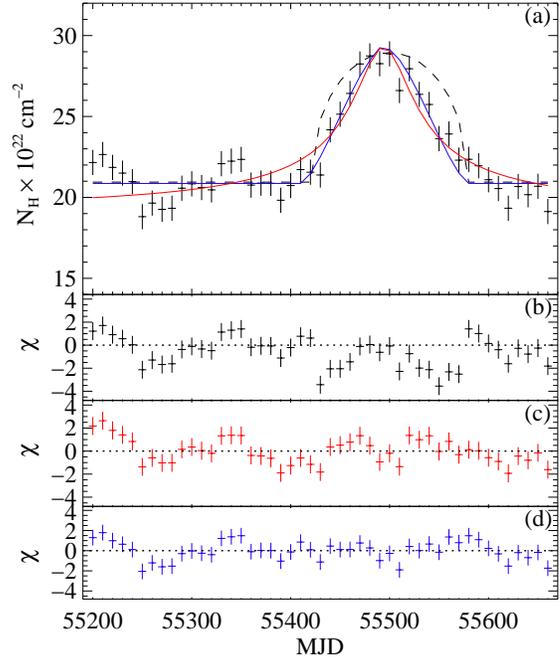}
  \caption{Three model fits to the \nh profile.  Panel (a) shows the data with three models: a uniform sphere (black dashed line), a $\beta$ model
  profile (blue line), and a linear-density sphere with maximum density at the center falling linearly with radius (red line).  Panel (b) shows fit 
  residuals to the uniform sphere model, panel (c) shows residuals to the $\beta$ model, and panel (d) shows residuals to the linear-density
  sphere model.}
  \label{fignhfit}
\end{figure}

We tested three density profiles for the increase in \nh above a constant baseline which was left free: a sphere of uniform density; 
a $\beta$ profile as used by Lamer \etal (2003) to fit a similar occultation observed in NGC~3227 given by the equation, 
\begin{center}
\begin{equation}  N_\text{H}(r) = N_\text{H center} \times \sqrt{1-(r/R_\text{c})^{2}}  \end{equation}
\end{center}
where R$_\text{c}$ is the core radius (Dapp \& Basu 2009); 
and a linear-density sphere with a maximum central density and a density profile described by the equation,
\begin{center}
\begin{equation}  \rho(r) = \rho_{\rm center} \times \frac{(R-r)} {R}  \end{equation}
\end{center}
where R is the outer radius of the spherical clump.  Figure \ref{fignhfit} shows the data, models and fit residuals for all three models.
The linear-density sphere gave the best fit with $\chi^2$/dof = 73/43.  The $\beta$ model gave  $\chi^2$/dof = 96/43
and the uniform sphere gave an unacceptable fit with $\chi^2$/dof = 160/43.
For the linear-density sphere we found that the occultation lasted a total of 170 days with a maximum column density of 
8.4 $\times 10^{22}$\,cm$^{-2}$ above a baseline of 20.9 $\times 10^{22}$\,cm$^{-2}$.  
The core radius crossing time (FWHM) for the $\beta$ model was 60 days with a maximum column density of 
10.3 $\times10^{22}$\,cm$^{-2}$ above a constant baseline level of absorption of 19.0 $\times10^{22}$\,cm$^{-2}$.
For analysis of the physical attributes of the clump based on these models see Section 3.

We also tested for cometary tails as seen by Maiolino \etal (2010), testing for asymmetry in the ingress and egress durations,
however these did not improve the fit.  From visual inspection it
is clear that the \cena occultation is fairly symmetrical with a smooth, gradual increase and decrease in the column density whereas
a comet-like shape would be appropriate for a rapid increase and slow decrease.  Consequently, this model is inappropriate for this source.


\begin{figure}
  \plotone{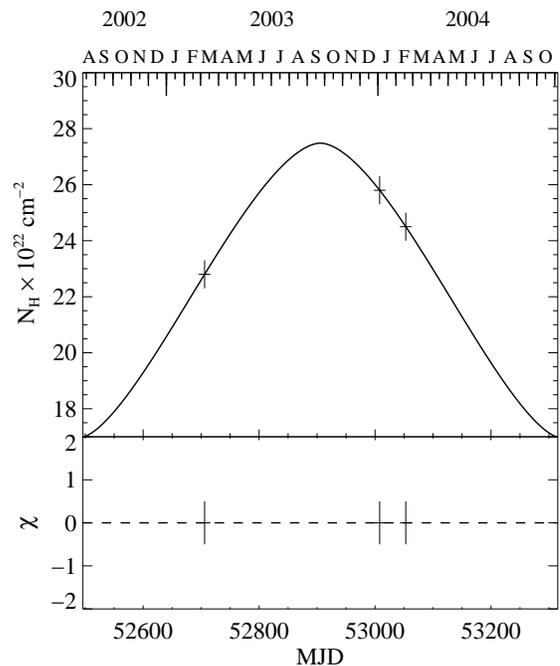}
  \caption{Fitting the linear-density sphere to the 2003--2004 increase in \nh.  The solid line represents a single clump occulting 
  the source with data--fit residuals shown in the lower panel.  }
  \label{figrlin}
\end{figure}


\section{Discussion}

Of the other AGN with similar short term increases or decreases in \nh, many
have posited that these clouds are part of the BLR rather than constituents of the torus. 
The low density of the clumps observed in \cena combined with the lack of detected 
broad lines in this source (Alexander \etal 1999) make this a very unlikely scenario.  
Therefore we conclude that the transiting clump(s) that have been seen in this source 
must arise from the dusty torus.  

The clumpy torus model of Nenkova (Nenkova \etal 2008a;  Nenkova \etal 2008b) predicts a small number of clumps 
along the line of sight to the nucleus making up the total observed column density.  This is borne out in our observation.  
The ratio of the baseline column density to the average increase caused by the clump is $\sim$ 5.  We can therefore 
assume that the number of clouds along the line of sight is $\lesssim$ 5 since some of the baseline absorption column 
may be dust free and reside inside the inner radius of the dusty torus.  For example, the very short transits seen by 
Risaliti \etal (2009) must come from inside the dust sublimation radius.  Assuming a viewing angle to the equatorial plane of 
62.$\degr$6 and an angular distribution of clouds of 60$\degr$ (Ramos Almeida \etal 2009)
we calculate that the average number of clouds along an equatorial ray is $\lesssim$ 14.  This is consistent with Rothschild \etal (2011)
as well as the predicted number of clumps from Nenkova \etal (2008b)  which is no more than 10--15 clumps along an equatorial ray.

Assuming that all of these clouds have an average column density of 4 \ttt this would imply a total column density along the equator
of $\sim\,$6 $\times 10^{23}$\,cm$^{-2}$ which is just on the verge of being Compton-thick.  This is reasonable given that no
Compton reflection signal has ever been detected in this source.

The distance of the dusty torus in \cena from the central illuminating source derived from infrared 
measurements by Meisenheimer \etal (2007) was 0.1--0.3 pc.  Nenkova \etal (2008b) calculated an inner dust sublimation radius
of 0.04 pc.  We can place limits on the radius of the torus by following the calculations of Lamer \etal (2003) which use the 
information that the obscuring material is completely cold with an ionization parameter $\lesssim\,$1 and assuming Keplerian motion.
Their equation (3) gives a relationship between the radius of the material and the ionizing luminosity ($L_\text{ion}$) of the source:
\begin{center}
\begin{equation} R \simeq 4 \times 10^{16}\, M_{7}^{1/5} \left ( \frac {L_{42}\, t_\text{days}}{N_{22}\, \xi}  \right )  \end{equation}
\end{center}

Where $M_{7}$ is the mass of the black hole which has been measured at M$_{BH} = 6 \times 10^7$ \Msun (Cappellari \etal 2009; Neumayer \etal 2010),
$L_{42} = L_\text{ion}/10^{42}$ where $L_\text{ion} \approx 3 \times10^{42}$ erg s$^{-1}$ is the ionizing radiation at 13.6 keV, 
$t_\text{days}$ is the crossing time of the event,
and $N_{22}$ is the maximum column density of the clump.  Assuming $\xi$=1 erg cm s$^{-1}$ gives a minimum distance to the torus of $\sim$0.1 pc, consistent
with the values found from infrared measurements.  We adopt an inner radius of 0.1 pc and an outer radius of 0.3 pc.

Using parameters determined by the linear-density sphere model (this model gives a better fit to the data than the other models, though it is
a purely empirical model) we can calculate the size of the clump and quantify the density profile.  
To begin we adopted the assumptions made in Rothschild et al. (2011) for Keplerian motion at 0.1--0.3 pc around the 6 $\times\,\,10^{7}$ 
M$_{BH}$ black hole and calculated a clump velocity of $\sim$930--1600 km/s.  
Combining this with the measured 170 day transit we found a linear-density sphere with a 
diameter of 1.4--2.4 $\times 10^{15}$\,cm with a central number density of n$_{\rm H}$ = 1.8--3.0 $\times 10^7$\,cm$^{-3}$.  
The total mass of this clump can be approximated as 2--5 $\times 10^{28}$ g or about 3--10 times the mass of the Earth.

For comparison, the occultation (if indeed it was a single absorption event) inferred by Rothschild \etal (2011) lasted between 
1 and 4 years with an increase in column density of 
$\sim\,$6 $\times 10^{22}$\,cm$^{-2}$.   Assuming a single uniform sphere it was found that the clump would have
an inferred length of 3--12 $\times 10^{15}$\,cm and a number density of 1--3  $\times 10^7$\,cm$^{-3}$.  

Fitting this occultation event with a single linear-density sphere model we found a length of 7$\times 10^{15}$\,cm and a central density of 
8 $\times\,\,10^6$\,cm$^{-3}$.  The best fit model with residuals is shown in Figure \ref{figrlin}; notice that with only three data points we needed
to assume the baseline column density rather than leave it as a free parameter as we did with the more recent event.  For this baseline
we have chosen the longterm average value given in Rivers \etal (2011) of 16.9 $\times 10^{22}$\,cm$^{-2}$.
This is somewhat larger and more diffuse than the more recent clump by a factor of 2--3, however this can easily be explained by 
inherent variation in clump sizes within the torus.  
Alternatively two or three much smaller, more dense clumps could adequately fit these three data points with similar characteristics 
to the more recent event, however with so few data points it is impossible to place constraints on such a scenario.

In conclusion, we have taken advantage of sustained monitoring by \xte to observe an occultation event in \cena in detail
from ingress to egress.
A discrete clump of material likely associated with a clumpy torus transited the line of sight to the central illuminating source 
for 170 days between 2010 August and 2011 February with a maximum increase in \nh of 8.4 $\times 10^{22}$\,cm$^{-2}$.
Assuming the clump of material was roughly spherical with a linear density profile and assuming a distance from the central nucleus of 
0.1--0.3 pc we found that the clump had a linear dimension of 1.4--2.4 $\times 10^{15}$\,cm with a central number density of 
n$_{\rm H}$ = 1.8--3.0 $\times 10^7$\,cm$^{-3}$, in good agreement with previous results.  Two occultation events seen in 
$\sim$10 years confirm that clumps of material are indeed transiting our line of sight and evidence suggests that they are
part of a clumpy, Compton-thin torus, the characteristics of which are consistent with the model proposed by Nenkova \etal (2008a; 2008b).

\begin{acknowledgments}
This research has made use of data obtained from the \textsl{RXTE} satellite, a NASA space mission.
This work has made use of HEASARC online services, supported by NASA/GSFC, and the NASA/IPAC Extragalactic Database, 
operated by JPL/California Institute of Technology under contract with NASA.
The research was supported by NASA Contract NAS 5-30720 and Grant NNX11AD07G.
\end{acknowledgments}


\end{document}